\documentclass[apj]{emulateapj}
\usepackage{natbib}
\usepackage{amsmath}
\usepackage{rotating}
\usepackage{enumerate}
\usepackage{apjfonts}

\begin{document}

\title{Observational implications of gamma-ray burst afterglow jet simulations and numerical light curve calculations}
\author{Hendrik J. van Eerten, Andrew I. MacFadyen}
\affil{
  Center for Cosmology and Particle Physics, Physics
  Department, New York University, New York, NY 10003}

\begin{abstract}
We discuss jet dynamics for narrow and wide gamma-ray burst (GRB) afterglow jets and the observational implications of numerical simulations of relativistic jets in two dimensions. We confirm earlier numerical results that sideways expansion of relativistic jets during the bulk of the afterglow emission phase is logarithmic in time and find that this also applies to narrow jets with half opening angle of 0.05 radians. As a result, afterglow jets remain highly nonspherical until after they have become nonrelativistic. Although sideways expansion steepens the afterglow light curve after the jet break, the jet edges becoming visible dominates the jet break, which means that the jet break is sensitive to the observer angle even for narrow jets. Failure to take the observer angle into account can lead to an overestimation of the jet energy by up to a factor 4. This weakens the challenge posed to the magneter energy limit by extreme events such as GRB090926A. Late time radio calorimetry based on a spherical nonrelativistic outflow model remains relevant when the observer is approximately on-axis and where differences of a few in flux level between the model and the simulation are acceptable. However, this does not imply sphericity of the outflow and therefore does not translate to high observer angles relevant to orphan afterglows. For more accurate calorimetry and in order to model significant late time features such as the rise of the counterjet, detailed jet simulations remain indispensable.
\end{abstract}

\section{Introduction}
\label{introduction_section}
Gamma-ray burst (GRB) afterglows are often theorized to result from synchrotron radiation from a decelerating relativistic blast wave (\citealt{Meszaros1997}, see also reviews by \citealt{Meszaros2006, Piran2005} and references therein). Over the past few years, different groups have performed two-dimensional (2D) relativistic hydrodynamics (RHD) GRB afterglow jet simulations at very high resolution, coupled to synchrotron radiation calculations (\citealt{Zhang2009, vanEerten2010c, vanEerten2011, Wygoda2011, vanEerten2011sgrbs}). The two main reasons to employ RHD simulations are that both jet decollimation and deceleration are very difficult to capture in detail in analytical models. While self-similar solutions exist for both the ultra-relativistic Blandford-McKee (BM) phase (\citealt{Blandford1976}, assuming a conic section of this spherical solution can be applied before the jet edges come into causal contact) and the late nonrelativistic Sedov-Taylor (ST) phase (\citealt{Taylor1950, Sedov1959}), no fully self-consistent model exists connecting both regimes. The recent numerical studies cited above have begun to systematically explore the transrelativistic behavior of afterglow jets. In this paper we address the following questions regarding jet dynamics and the shape of light curves calculated directly from recent high resolution numerical simulations:

\begin{enumerate}
 \item Recent numerical studies (\citealt{Zhang2009,Wygoda2011}) agree on the absence of exponential sideways expansion of the jet (as predicted by \citealt{Rhoads1999} to occur in the relativistic regime) for jets with a reasonably wide half opening angle of 0.2 radians.  Is this type of expansion significant for lateral expansion of narrower jets?
 \item Is the logarithmic spreading (\citealt{Zhang2009}) common to both narrow and wide jets and what does that imply for transition to the non-relativistic regime?
 \item What do the answers to question 1 and 2 imply for the shape of the light curve as set by the competing effects of jet spreading and the jet edges becoming visible due to a decrease in relativistic beaming of the radiation? 
 \item Will the observer angle still play a role for the observed jet break of narrow jets, as it does for wide jets?
 \item What are the implications for late time radio observations, especially with respect to calorimetric estimates of the jet energy?
\end{enumerate}
Related to question 4 is the issue of the importance of the observer angle for the energy estimate of the GRB from the jet break. It has been claimed in observational studies \citep{Cenko2010, Cenko2011} that the magnetar model for GRB progenitors is severely challenged by recent observations (e.g. GRB 090926A). In view of this, we will also return to the question
\begin{enumerate}
 \setcounter{enumi}{5}
 \item Is the magnetar model still challenged when the observer angle is taken into account in determining the jet energy from the jet break?
\end{enumerate}
that we briefly discussed in \cite{vanEerten2010c}.

The rest of this paper is organized as follows. After a brief interlude establishing some notation and relevant timescales, in section \ref{jet_dynamics_section} we discuss the dynamics of afterglow jets (i.e. questions 1, 2). In section \ref{observational_implications_section} we discuss the observational implications (questions 3-6). We summarize our results in section \ref{conclusions_section}.

\section{Timescales}
Analytical studies have argued for a number of timescales of special relevance. Some of these timescales will feature in the discussion below and we therefore define them here.

The first of these is the point in time that approximately marks the onset of lateral spreading of the jet. During the BM phase, and before any significant lateral spreading, the radial outflow decelerates according to $\gamma = \gamma_0 (t / t_0)^{-3/2}$, with $\gamma$ the Lorentz factor of the fluid directly behind the shock. Here, $\gamma_0$ and $t_0$ are determined completely by the explosion energy and circumburst medium density, the physical parameters that set the scale of a BM point explosion. When the flow has decelerated to the point where $\gamma \sim 1 / \theta_0$, where $\theta_0$ the original jet half opening angle in radians, sideways expansion is expected to take off. Two motivations for this are that this time $t_\theta$ marks approximately the time when the edges of the jet achieve causal contact \citep{Shapiro1979}, as well as the point where the widening of the jet becomes comparable to the original jet opening angle (under the assumption that the relativistic jet spreads with the speed of sound, \citealt{Rhoads1999}). Using conservation of energy in the expanding blast wave (eq. 43, BM), we find
\begin{equation}
 t_\theta = 235 (E_{iso} / 10^{53} n_0)^{1/3} (\theta_0 / 0.1)^{2/3} \textrm{ days}.
\end{equation}
Here $E_{iso}$ denotes the isotropic equivalent energy of the jet in ergs and $n_0$ the circumburst medium number density in cm$^{-3}$. Aside from the constant light travel time from the origin of the explosion to the observer, the observed time $t_{\theta, \oplus}$ differs significantly from the time in the progenitor frame because the jet almost keeps up with its own radiation. For the front of the jet at radius $R$, this transformation implies
\begin{equation}
 t_{\theta,\oplus} = t_\theta - R(t_\theta) / c = \frac{t_\theta}{16 \gamma^2},
\end{equation}
where we have used $R = c t (1 - 1 / 16 \gamma^2)$ (eq 26, BM). Here $c$ denotes the speed of light and $\gamma$ refers to the fluid Lorentz factor directly behind the shock front.

The outflow has been argued \citep{Livio2000} to become roughly spherical by the time $t_{s,\oplus}$ (in the observer frame), expressed as
\begin{equation}
t_{s,\oplus} = 230 \left( \frac{E_{iso}}{10^{53} n_0} \right)^{1/3} \left( \frac{\theta_0}{0.1} \right)^{2/3} \textrm{ days},
\label{tsobs_equation}
\end{equation}
in \cite{Wygoda2011}. Here $E_{iso}$ denotes the isotropic equivalent energy of the jet in ergs and $n_0$ the circumburst medium number density in cm$^{-3}$. If $t_{s,\oplus}$ marks the point where the jet becomes quasi-spherical, a direct implication is that beyond this point in time the original orientation of the jet with respect to the observer has become irrelevant. Note that $t_{s, \oplus} \approx t_\theta$ (the difference between 235 and 230 being a difference in round-off), even though the two timescales are expressed in different frames. The implication is therefore that the difference between frames has become neglible by this point in time. This is because two assumptions have been made in the derivation of $t_{s, \oplus}$: it is assumed that lateral spreading is a very fast process that takes approximately $t_\theta \sim R(t_\theta) / c$ to complete, and that during the spreading phase further increase in radius compared to $R(t_\theta)$ is neglible. Later we will see that the second of these assumptions leads to a grossly overestimated $t_{s, \oplus}$ compared to what fast spreading would imply, and that as a result $t_{s,\oplus}$ in practice becomes comparable to the time when the onset of the nonrelativistic phase is observed.

After this time $t_{s,\oplus}$, the jet is expected \citep{Piran2005} to further decelerate and become nonrelativistic at time
\begin{equation}
t_\textrm{NR} = 1100 \left( \frac{E_{iso}}{10^{53}n_0} \right)^{1/3} \textrm{ days}.
\end{equation}
It is worth noting that $t_{s,\oplus}$ depends on the original jet opening angle, whereas $t_\textrm{NR}$ does not. We also note that the numerical factor in the equation for $t_{s,\oplus}$ given in \cite{Zhang2009} is approximately twice that in \cite{Wygoda2011} and the numerical factor in \cite{Zhang2009} for $t_\textrm{NR}$ is 970 rather than 1100 days. We follow \cite{Wygoda2011} in order to compare directly to their work, but these differences do not alter our conclusions. Because at $t_\textrm{NR}$ the jet has become nonrelativistic, it is assumed that $c t_\textrm{NR} \gg R(t_\textrm{NR})$ and that as a result $t_\textrm{NR} \approx t_{\textrm{NR}, \oplus}$.

\section{Jet dynamics}
\label{jet_dynamics_section}

We have used the RHD code \textsc{ram} \citep{Zhang2006} to run a number of high-resolution jet simulations starting from the BM solution. A series of jets with $\theta_0 = 0.05$ rad and a wide jet with $\theta_0 = 0.2$ rad, starting from fluid Lorentz factors\ $\gamma = 40$ and $\gamma = 25$ at the shock front
\footnote{Simulations with initial $\gamma > 1 / \theta_0$ are
  expected to lead to nearly identical fluid evolution (and light
  curves), as lateral spreading has not yet begun and the radial
  outflow is still expected to follow BM as confirmed by our simulations starting with
  different Lorentz factors} 
and both with $E_{iso} = 10^{53}$ have been
calculated as well as a $\theta_0 = 0.05$ rad jet with an energy in
both jets of $E_j = 2 \times 10^{51}$ erg ($E_{iso} = 2 E_j /
\theta_0^2 = 1.6 \times 10^{54}$ erg). In all cases, $n_0 = 1$. The
simulation resolution in all cases is similar to \cite{Zhang2009} and the
same adaptive mesh refinement strategies as described in
\cite{vanEerten2011sgrbs} have been applied. 

Our simulations use an EOS which smoothly interpolates between
relativistic ($\Gamma_{ad} = 4/3$) and non-relativistic ($\Gamma_{ad}
= 5/3$) \citep{Mignone2005}. The effect of the EOS on the dynamics and
the shape of the light curve are discussed in detail in
\cite{vanEerten2010}. There it was found (for spherical outflow) that
keeping $\Gamma_{ad} = 4/3$ led to a 9 percent smaller radius at late
time compared to a changing $\Gamma_{ad}$. The interpolating EOS will
lead to steeper light curves in the transrelativistic phase.

\begin{figure}
 \centering
 \includegraphics[width=0.9\columnwidth]{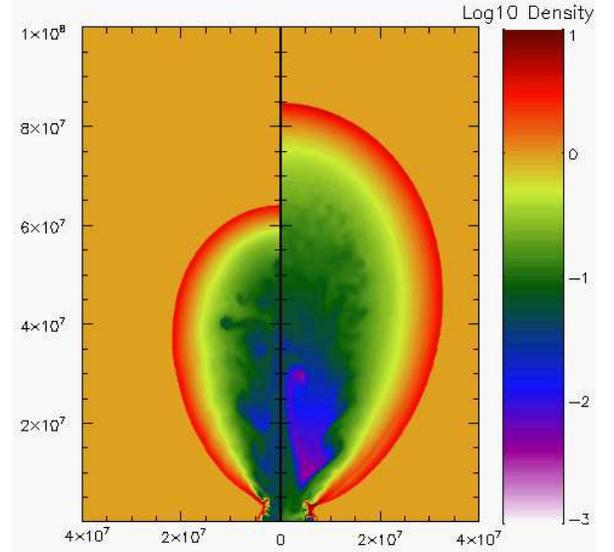}
 \caption{A snapshot of the comoving number density at $t = 0.95 \times t_\textrm{NR} \sim 9 \times 10^7$ s for the $\theta_0 = 0.05$ rad, $E_j = 1.25 \times 10^{50}$ erg simulation (left) and the $\theta_0 = 0.2$ rad, $E_j = 2 \times 10^{51}$ erg simulation (right). Both simulations have the same $E_\textrm{iso}$. The distances are in lightseconds. The outer radius along the axis of the narrow jet is $1.9 \times 10^{18}$ cm and $2.55 \times 10^{18}$ cm for the wide jet, while $c \times t = 2.7 \times 10^{18}$ cm. }
 \label{density_figure}
\end{figure}

Fig. \ref{density_figure} compares between narrow and wide jet at
$0.95 t_\textrm{NR} \approx 10^3$ days. The narrow jet has $E_j = 1.25
\times 10^{50}$ erg, the wide jet $E_j = 2 \times 10^{51}$ erg
(matching Fig. 2, \cite{Wygoda2011}). The different $E_j$ values lead to different
radii around $t_\textrm{NR}$. When we compare our
Fig. \ref{density_figure} to Fig. 2 in \cite{Wygoda2011}, we find excellent
agreement in the width and overall radial extent of the jets in \cite{Wygoda2011}
and our simulations.

We caution against interpreting the differences in radius for the wide
and narrow jet visible in Fig. \ref{density_figure} as proof of fast
early time spreading and deceleration of the narrow jet relative to
the wide jet. Noticeable differences in radius between narrow and wide
jet occur after $\gamma \sim 1$ by definition and could therefore be
attributed to $E_j$ rather than early time spreading or $\theta_0$. At
this stage $E_\textrm{iso}$ no longer applies due to causal contact
across all angles of the jet and neither does the assumption of ST
self-similarity as long as the jet is not spherical and an additional
length scale is introduced by the current width of the jet.

\begin{figure}
 \centering
 \includegraphics[width=0.9\columnwidth]{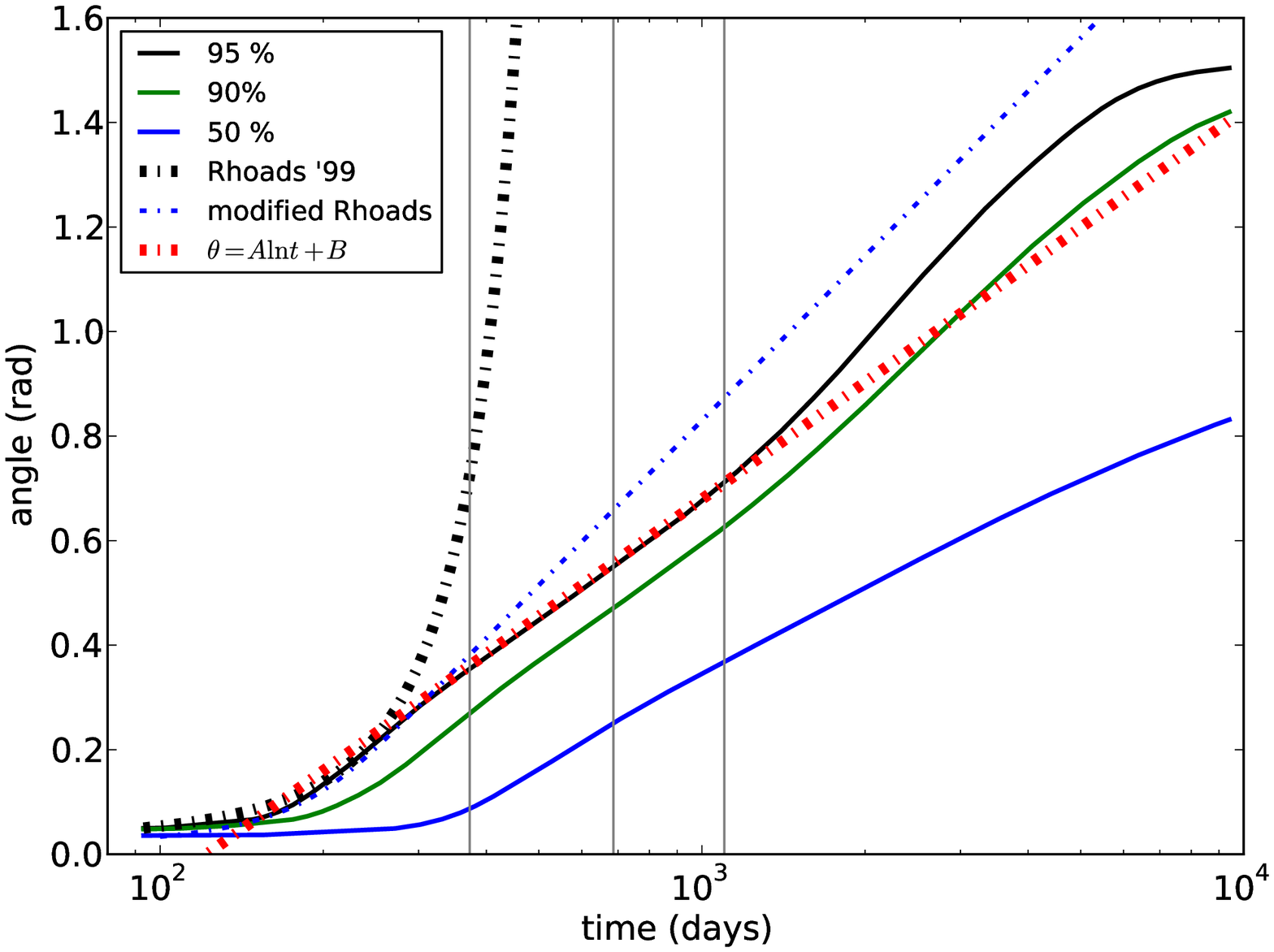}
 \includegraphics[width=0.9\columnwidth]{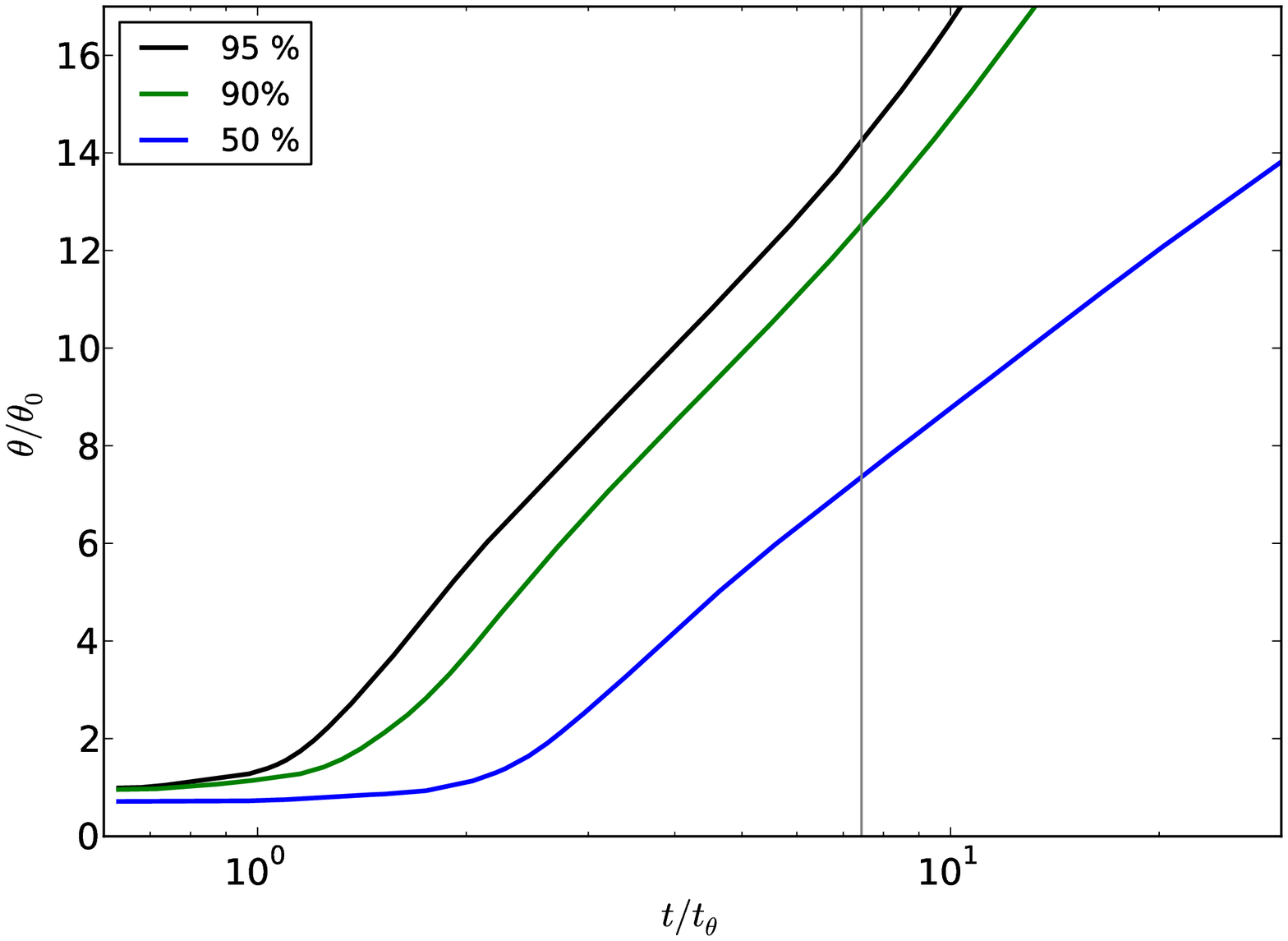}
 \caption{Top: Outer angle of three regions containing fixed fractions of total jet energy for the $\theta_0 = 0.05$ rad, $E_j = 1.25 \times 10^{50}$ erg jet. The vertical grey lines denote $t(\gamma = 5)$, $t(\gamma = 2)$ and $t_\textrm{NR}$ from left to right. The dash-dotted black curve denotes exponential expansion using $\theta = \theta_0 \exp [ c (t - t_0) / l_\textrm{SNT} ]$, the dash-dotted blue curve the modified Rhoads model fit to $t(\gamma=5)$ and the dash-dotted red curve a logarithmic expansion curve fit to the 95 \% curve between the onset of spreading and $t_\textrm{NR}$. Bottom: same as top, only rescaled using $\theta_0 = 0.05$ and $t_\theta = t(\gamma=1/\theta) \sim 148$ days, for direct comparison with Fig. 1 of \cite{Wygoda2011}. The single vertical grey line denotes the scaled $t_\textrm{NR}$.}
 \label{theta_spreading_figure}
\end{figure}

We now examine the angular distribution of jet energy for the narrow jet with $E_j = 1.25 \times 10^{50}$ erg. In Fig. \ref{theta_spreading_figure} we have plotted the evolution of the boundaries marking the regions containing different fractions of the total jet energy. Before discussing the behavior of the 95 \% boundary we emphasize that its interpretation as `edge' of the jet will become ambiguous if the front of the jet is not homogeneous (i.e. when energy and Lorentz factor depend on angle with respect to the jet axis). This inhomogeneity is a key finding of previous numerical work \citep{Zhang2009, vanEerten2011, Granot2001, Granot2007}.

Taking the 95 \% boundary at face value and comparing it with exponential expansion, we see from Fig. \ref{theta_spreading_figure} that both curves start to deviate before $\gamma \sim 5$ (for exponential expansion, $t = t_0 \exp [c(t-t_0)/l_{SNT}]$, where $t_0$ denotes the time when $\gamma = 1 / \theta_0$, $c$ the speed of light and $l_{SNT} \approx 2.7 \times 10^{17}$ cm the Sedov length for $\theta_0 = 0.05$ rad., see \cite{Zhang2009}). The 95 \% opening angle at $\gamma = 5$ is $\sim 7 \theta_0 \sim 0.35$ rad is large with respect to $\theta_0$, but still only a small fraction of the final opening angle, meaning that most lateral expansion will not take place in the strongly relativistic regime.

The exponential curve can be improved upon by not taking the exponential limit when implementing the dynamical equations from \citealt{Rhoads1999}. We obtain the dash-dotted blue curve shown in Fig. \ref{theta_spreading_figure} when following \cite{Wygoda2011} by substituting $\textrm{d}\theta(r)/r \propto 1 / \gamma r$ for eq. (3) in \cite{Rhoads1999} and including an additional scaling factor determined by fitting to the simulation curve. However, \emph{logarithmic} expansion provides a better fit, not only beyond $\gamma \sim 2$ or 5, but for the entire region up to $t_\textrm{NR}$, even for narrow jets. The values for this fit function have been determined from $\theta$ at $t_\textrm{NR}$ and $\theta$ at the onset of the sideways movement of the 95 \% boundary ($t \sim 150$ days). Modified Rhoads ends in a logarithmic increase but is too constrained to get the slope right.

Finally, Fig. \ref{theta_spreading_figure} shows that the 95 \% boundary lies at 0.7 rad for the narrow jet at $t_\textrm{NR}$, but an extended phase of lateral expansion still follows during which the increase of the curve even temporarily becomes steeper than at earlier times. The non-relativistic dynamics depend on $E_j$ rather than $E_{iso}$ and $E_j$ jet rather than $\theta_0$ determines the late-time sideways expansion. In the bottom plot the angles are rescaled as fraction of the original opening angle and the time as fraction of the time when $\gamma = 1 / \theta_0$, for comparison with Fig. 1. in \cite{Wygoda2011}. Note especially that the 90 and 50 \% boundaries show \emph{no} lateral spreading at $\gamma = 1 / \theta_0$

\begin{figure}
 \centering
 \includegraphics[width=0.9\columnwidth]{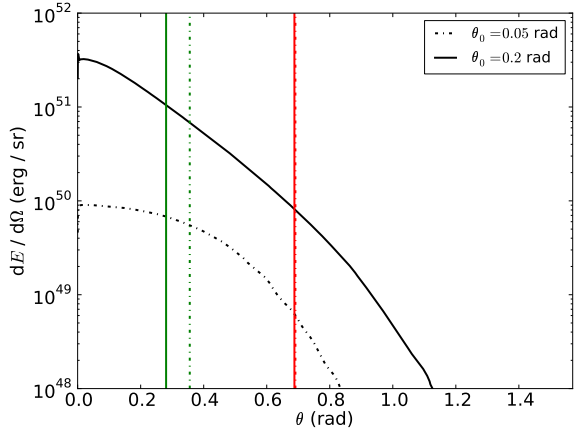}
 \caption{Angular distribution of energy at $t = 0.95 t_\textrm{NR}$ for both wide (0.2 rad, solid black curve) and narrow (0.05 rad, dash-dotted black curve) jet. The two vertical green lines on the left denote the outer angle of the region containing 50 \% of the jet energy (again, solid for 0.2 rad, dash-dotted for 0.05 rad). The two red lines on the right denoting the outer angle of the region containing 95 \% are nearly indistinguishable.}
 \label{angle_distro_figure}
\end{figure}

However, as stated above, arguments invoking the 95 \% boundary become less meaningfull if the shock front is inhomogeneous. Fig. \ref{angle_distro_figure} shows that at $t = 0.95 t_\textrm{NR}$ (i.e. the same time as in our Fig. \ref{density_figure} and Fig. 2 of \cite{Wygoda2011}) there is still an order of magnitude difference in energy between the 50 \% and the 95 \% regions, directly confirming that the shock fronts remain inhomogeneous even on very long timescales. On the other hand, the 95 \% boundaries for the wide and narrow jets lie very closely together at this point in time, suggesting that at least superficially the jet retains less memory of $\theta_0$ and $E_j$ is becoming the dominant factor determining jet dynamics.

We conclude that even for narrow jets no phase of exponential spreading occurs that leads to approximate sphericity over only a logarithmic increase in radius, as argued by \cite{Livio2000} (note that $t_s \ll t_\textrm{NR}$): Exponential expansion does not fit the light curve down to $\gamma \sim 5$ for $95 \%$ and the $50 \%$ boundary only starts spreading in the transrelativistic regime. Instead, the expansion accross the entire transrelativistic regime is logarithmic, as described in \cite{Zhang2009} (and as analytically expected, see the appendix of \citealt{vanEerten2010c}). We find that the jets remain strongly inhomogeneous and highly nonspherical at late times.

\section{Observational implications}
\label{observational_implications_section}

\subsection{jet break and observer angle}

\begin{figure}
 \centering
 \includegraphics[width=0.9\columnwidth]{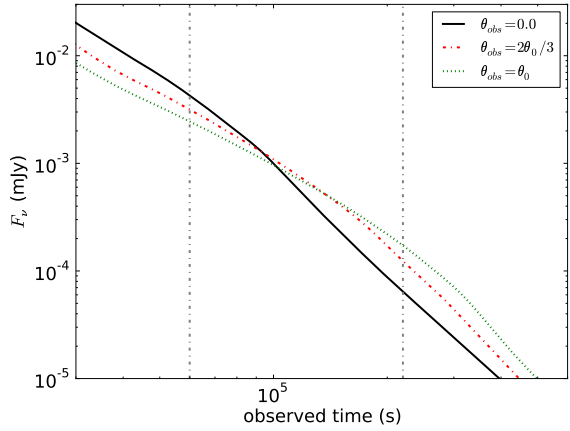}
 \caption{X-ray afterglow light curves showing the jet break for the narrow 0.05 jet with $E_j = 2\times10^{52}$ erg typical to \emph{Swift} light curves, for observers at different angles. The calculation uses $z=0.5$, luminosity distance $d_L = 8.7 \times 10^{27}$ cm, synchrotron slope $p = 2.5$ and the method from \cite{Zhang2009}. Vertical grey lines denote jet break fit results for $\theta_{obs} = 0, \theta_0$.}
 \label{swift_light_curve_figure}
\end{figure}

Having established that even for narrow jets the jet dynamics are not dominated by exponential expansion, it follows that the observed jet break is also likely not dominated by lateral expansion of the flow. Even for narrow jets, features associated missing flux due to the jet edges becoming visible are still expected to play a role in, if not dominate, the shape of the light curve. One such feature is the role of the observer angle (\citealt{vanEerten2011, vanEerten2011sgrbs, vanEerten2011proceedings}). Fig. \ref{swift_light_curve_figure} shows X-ray light curves (1.5 KeV) for a narrow jet with $E_j = 2 \times 10^{52}$ erg seen from different observer angles. Given that the observed jet break time $t_j$ was found to be set by the edge furthest from the observer, \cite{vanEerten2010} suggest
\begin{equation}
 t_j = 3.5 (1+z) \left( \frac{E_{iso}}{10^{53}} \right)^{1/3} n_0^{-1/3} \left(\frac{\theta_0 + \theta_{obs}}{0.2}\right)^{8/3} \textrm{ days},
\label{jetbreak_time_equation}
\end{equation}
where $z$ denotes redshift. In practice jet spreading, radial fluid structure and arrival time effects render the light curve more complex than mere power laws. Smooth power law fits assuming a fractional error of 10 pct on the datasets from Fig. \ref{swift_light_curve_figure} lead to a jet break time $6.0 \times 10^4 \pm 4 \times 10^3$ s on-axis ($\chi^2 / dof = 0.25$) and $2.22 \times 10^5 \pm 6 \times 10^3$ s on-edge ($\chi^2 / dof = 0.26$). 

Although the observer angle effect may therefore not be as severe as implied by eq. \ref{jetbreak_time_equation}, this does confirm that the error due to observer angle in energy estimates based on jet opening angles is a general issue that not only applies to wide jets. In \cite{Cenko2011} GRB 090926A, with prompt energy release $E_\gamma = 1.4 \times 10^{52}$ erg, is found to exhibit the most severe challenge to the magnetar energy limit of $3 \times 10^{52}$ erg. To illustrate: downscaling the break time by a factor 3.7, downscales the jet opening angle by a factor 1.6 and $E_\gamma$ by a factor 2.7, leading to $E_\gamma \sim 5.2 \times 10^{51}$ erg. Although we therefore conclude that the challenge posed to the magnetar model by events like GRB 090926A might be weaker than previously reported, we emphasize that we draw no conclusions regarding the validity of the magnetar model and that the significant finding from \cite{Cenko2011} that a class of very energetic (relative to earlier observations) GRB events is emerging remains unaltered.

If jet breaks for narrow and wide jets exhibit the same general shape, earlier conclusions regarding the effect of jet spreading for non-narrow jets are general as well. Specifically, this implies that, while subdominant, jet spreading \emph{does} affect the post-break slope. This effect has been demonstrated explicitly in \cite{vanEerten2011} and earlier in \cite{Granot2007}. We note that jet expansion (even if logarithmic) can not be fully ignored when making quantitative predictions \cite{Zhang2009}.

\subsection{Late time calorimetery and $t_s$}

Given the good agreement between the recent wide jet simulations \citealt{Zhang2009,Wygoda2011}and that at late times $t > t_\textrm{NR}$ the shape of the light curve is determined by $E_j$ rather than $\theta_0$, we also support the conclusion from \cite{Wygoda2011} that using a Sedov-Taylor approximation based on $E_j$ to model the light curve at late times is a valid approach. For observers approximately on-axis, and when errors on the flux of the order of a few are acceptable, this approach is even viable at times $t > t_{s,\oplus}$, as defined by \cite{Wygoda2011} (e.g. for the analysis in \citealt{Frail2000}). But this does \emph{not} mean that the jet is approximately spherical by this time, and \emph{neither} does it confirm the assumption in \cite{Livio2000} of fast deceleration over a neglible increase in distance.

\begin{figure}
 \centering
 \includegraphics[width=0.9\columnwidth]{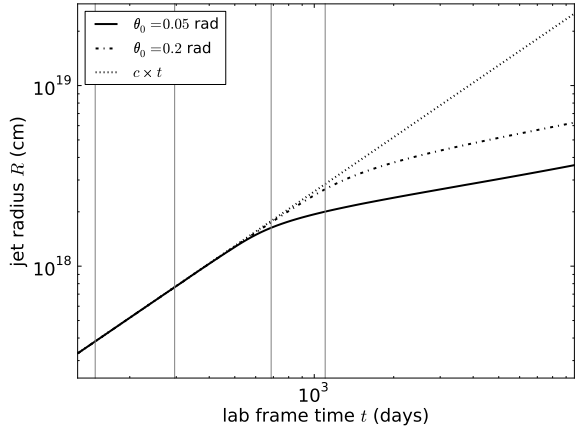}
 \caption{On-axis blast wave radius as a function of lab frame time for narrow and wide jets. The vertical lines denote from left to right: $t_\theta$ (onset of spreading), $t_s \sim 2 t_\theta$ (spreading originally expected to be completed), $t(\gamma=2)$ (BM approximation ceases to be applicable, deceleration expected to start to become apparent even for spherical jets), $t_{NR}$ (jet in non-relativistic regime).}
 \label{radialincrease_figure}
\end{figure}

The latter is illustrated in Fig \ref{radialincrease_figure}. It shows the on-axis blast wave radius both for the narrow and wide jet simulation. According to \cite{Wygoda2011}, $t_{s,\oplus}$ is given by eq. \ref{tsobs_equation}, where the increase in radius after $t_\theta$ is neglected. As a result, $t_{s,\oplus} \approx 148$ days for the narrow ($\theta_0 = 0.05$ rad) jet. However, if $t_s < t_{NR}$ the increase in radius can not be neglected compared to the increase in time since $t_\theta$, because $\gamma > 1$ implies that $R(t_s) \equiv R_s$ will be comparable to $c t_s$. From Fig \ref{radialincrease_figure}, we find $t'_{s,\oplus} = t_s - R'_s = 295.86 - 7.64455 \cdot 10^{17}$ cm$/c = 0.71$ days, where primed quantities are based on the simulation results rather than analytically estimated. For comparison, if there were no additional deceleration due to sideways spreading \emph{at all}, and the blast wave had kept on decelerating  according to $\gamma = \gamma_\theta (t / t_\theta)^{-3/2}$ (BM), the resulting $t''_{s,\oplus} = 0.37$ days. Clearly, 0.71 days lies closer to 0.31 days than to 148 days. For the wide blast wave $t'_{s,\oplus} = 15.8$ days, while $t_{s,\oplus} = 372$ days.

\cite{Wygoda2011} show that beyond $t_{s,\oplus}$, the difference between the simulation light curve and a simple non-relativistic model is no more than a factor of a few. And indeed, although the above demonstrates that it is not valid to interpret this as confirming the (modified) Rhoads model of quick relativistic spreading and deceleration, $t_{s,\oplus}$ remains useful as a rule-of-thumb to indicate the point when the non-relativistic model starts to become appropiate for calorimetry estimates (but depending on the amount of error that is deemed acceptable and only for observers close to the jet axis, see below). Mathematically compensating the lack of quick spreading by neglecting the increasing radius has the effect that $t_{s,\oplus}$ is put far into the transrelativistic regime (even for spherical blast waves) and as such it has the advantage of marking the turnover into the semi-spherical non-relativistic phase, while improving upon $t_{NR}$ by retaining a dependency on $\theta_j$. Insofar as a clear interpretation of $t_{s,\oplus}$ is possible, it is that $t_{s,\oplus}$ reveals the effects of $E_j$ and logarithmic spreading throughout the transrelativistic regime. In a practical sense, studies that depend on $t_{s,\oplus}$ remain valid for rough calorimetry estimates (e.g. \citealt{Frail2000, Shivvers2011}).

\begin{figure}
 \centering
 \includegraphics[width=0.9\columnwidth]{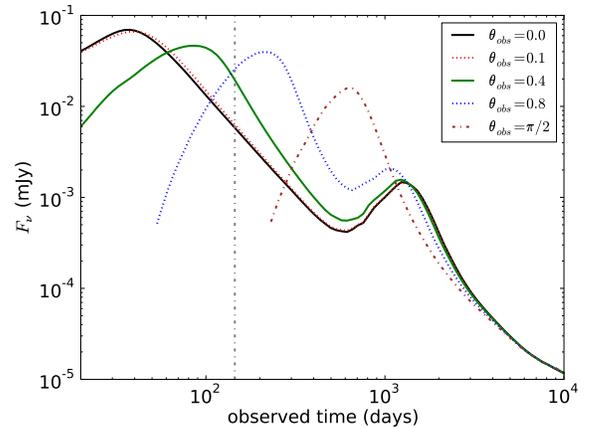}
 \caption{Late time radio light curves observed at 1.43 GHz, for a narrow jet with $\theta_0 = 0.05$ rad and $E_j = 1.25 \times 10^{50}$ erg. Synchrotron self-absorption is included but hardly affects the light curves. Light curves are plotted for various observer angles, including $\theta_\textrm{obs} > \theta_0$ (i.e. an orphan afterglow). The vertical dash-dotted grey curve denotes $t_s$, the point at which the jet was originally expected to become quasi-spherical. The calculation uses $z=0.5$, luminosity distance $d_L = 8.7 \times 10^{27}$ cm, $p = 2.5$ and the method from \cite{vanEerten2011sgrbs}.}
 \label{late_radio_light_curve_figure}
\end{figure}

For more accurate calorimetry and in order to model significant late time features such as the rise of the counterjet, detailed jet simulations remain indispensable. A straightforward illustration of nonsphericity is given in Fig. \ref{late_radio_light_curve_figure}, which shows radio light curves at $t > t_s$ for observers at high angles. For such \emph{orphan afterglows}, a calorimetry calculation based on the assumption of (semi-) sphericity beyond $t_s$ is severely limited given that the light curves can still differ between observer angles by more than an order of magnitude, even if the rapid variations may render orphan afterglows easier to detect by surveys. For observers slightly off-axis the issue is less problematic, as shown by Fig. \ref{late_radio_light_curve_figure}, and again, in general, one can assume that as long as the prompt emission is observed, radio calorimetry will yield approximately correct results. 

It should be noted however, that this does not extend to any modeling that includes a measurement of the \emph{slope} of the late time radio light curve. Even without effects such as the rise of counterjet (that even leads to a temporary increase in flux for a narrow jet), the light curve slope has not yet settled into the ST regime. This implies that an attempt to determine circumburst density parameter $k$ (for a profile $\rho \propto r^{-k}$) and/or the synchrotron slope $p$ from the light curve slope will still be bound by the late $5 t_{NR}$ value \citep{Zhang2009}, rather than $t_{s,\oplus}$. Similarly, a flattening of the light curve beyond $t_{s,\oplus}$ does not automatically imply the rise of a flux contribution from the host galaxy (e.g. as assumed for GRB 980703, \citealt{Berger2001}).

\section{conclusions}
\label{conclusions_section}

We draw the following conclusions with regard to the questions posed in section \ref{introduction_section}.
\begin{enumerate}
 \item There is no regime of exponential sideways expansion that noticably advances the onset of sphericity, not even for narrow jets. The front of the blast wave is far from homogeneous, which is a key requirement for the approximation by \cite{Rhoads1999} to hold. Even as material at the edges of the jet moves sideways, the bulk of the material remains unaffected for a long time.
 \item Sideways expansion of regions of fixed jet energy fraction is logarithmic in the entire region after the onset of expansion (different for different energy fractions) and $t_\textrm{NR}$. The jet remains highly unspherical at $t_\textrm{NR}$. The full transition to sphericity is a very slow process taking $\sim 5 t_\textrm{NR}$ to complete \citep{Zhang2009}.
 \item The post jet break light curve slope is determined by both spreading and jet edges becoming visible. Completely ignoring jet spreading at this stage will noticeably underestimate the steepening of the jet.
 \item Although its impact is slightly decreased by jet spreading, the observer angle remains important, even for narrow jets and off-axis observation of the jet will result in a light curve where the break time can be significantly postponed and even buried in the noise.
 \item Calorimetry based on applying a spherical nonrelativistic jet model to radio afterglows will remain relevant and approximately correct beyond $t_{s,\oplus}$, when the observer is approximately on-axis and errors of a few in flux level are considered acceptable. However, this does not imply sphericity of the outflow and therefore does not translate to high observer angles relevant to orphan afterglows. Also, $t_{s,\oplus}$ represents the effects of $E_j$ and long-term spreading during the transrelativistic phase, rather than quick early time spreading and deceleration (because the original derivation of $t_{s,\oplus}$ based on the latter underestimates the increase in radius during the spreading phase). For more accurate calorimetry and in order to model significant late time features such as the rise of the counterjet and features that depend on the light curve slope, detailed jet simulations remain indispensable.
 \item The magnetar model is challenged less by recent observations once allowance is made for the fact that the observer angle is not known when jet opening angle and energy are inferred from the jet break time. The inferred energy decreases by a factor 2-4.
\end{enumerate}

\section{Acknowledgements}
This research was supported in part by NASA through grant NNX10AF62G issued through the Astrophysics Theory Program and by the NSF through grant AST-1009863. Resources supporting this work were provided by the NASA High-End Computing (HEC) Program through the NASA Advanced Supercomputing (NAS) Division at Ames Research Center. The software used in this work was in part developed by the DOE-supported ASCI/Alliance Center for Astrophysical Thermonuclear Flashes at the University of Chicago. We wish to thank Weiqun Zhang, Nahliel Wygoda and Maxim Lyutikov for helpful discussion.\\

\end{document}